\documentclass[twoside,12pt,a4paper]{article}
\usepackage[dvips]{epsfig}
\usepackage{a4}
\usepackage{latexsym}
\usepackage{graphics}
\def\beq{\begin{equation}}
\def\eeq{\end{equation}}
\def\beqn{\begin{eqnarray}}
\def\eeqn{\end{eqnarray}}

\begin{document}

\title{Introduction: GDH and related topics}

\author{M.M. Giannini \\
{\small Dipartimento di Fisica and INFN, 
via Dodecaneso 33, 
I-16146 Genova, Italy}\\
{\small E-mail: giannini@ge.infn.it}}


\maketitle

\abstract{
The formulation of the Gerasimov-Drell-Hearn sum rule is revisited, showing its
connection with other sum rules occurring in electron scattering and discussing the
problem of its saturation, both from the theoretical and experimental point of view.
The generalisation to other nuclear targets is also reported.}

\section{Introduction}
This Conference is the second of the series, following the Mainz edition of 
two years ago \cite{mz}, testifying the growing interest of the hadronic physics 
community. Actually, the GDH sum rule is considered an important test of 
our knowledge of the electromagnetic excitation of the nucleon, since it is 
based on very general (and accepted) principles. On the other hand, the 
interest in the GDH has also been the starting point for a large variety of
activity, both experimental and theoretical, concerning, among others, the
electromagnetic  production of meson, the excitation of baryon resonances, 
the spin structure of the nucleon and of various nuclear targets.
In the last Conference, we have seen many interesting results, some of them,
specially the experimental ones, still at a preliminary level. In particular 
a big effort is devoted to the experimental verification of the saturation 
of the GDH, work which is now feasible thanks to the modern facilities. We 
all expect that on these (and many other) topics,  the Conference will produce 
exciting results.

In this introductory talk, I will briefly review the derivation of the GDH 
sum rule and point out some of the results presented two years ago for which 
some improvements and developments are expected.

\section{An introduction to GDH}

Here the main assumptions leading to the GDH sum rule are
briefly reported \cite{dd}.

The starting point is the consideration of the nucleon Compton scattering
amplitude $T(\nu,\theta)$, where $\nu$ is the photon energy and $\theta$
is the scattering angle. In the forward direction, the amplitude
$T(\nu,0)~=~T(\nu)$ can be written as
\begin{equation}
T(\nu)~=~{\vec{\epsilon}}_f^{~*} \cdot {\vec{\epsilon}}_i~f(\nu)~+
~i~({\vec {\epsilon}}_f^{~*} \times {\vec \epsilon}_i) ~\cdot
\vec{\sigma}~g(\nu),
\label{eq:fa}
\end{equation}
where ${\vec{\epsilon}}_f$ and ${\vec {\epsilon}}_i$ are
respectively the final and initial photon polarisation vectors, $\vec{\sigma}$
is the nucleon spin and $f(\nu)$ and $g(\nu)$ the non-spin-flip and the
spin-flip amplitudes, respectively. According to crossing symmetry
\begin{equation}
T(-~\nu,-~\gamma_f,-~\gamma_i)~=~T(\nu,\gamma_i,\gamma_f).
\label{eq:cross}
\end{equation}
This means that $f(\nu)$ ($g(\nu)$) is even (odd) in $\nu$.

For small values of the photon energy, the amplitudes can be expanded in a
power series. The first terms are fixed by the Low Energy Theorem (LET),
which is a consequence of Lorentz and gauge invariance \cite{lgg}. For the even
amplitude $f(\nu)$ we have:
\begin{equation}
4 \pi~f(\nu)~=~-\frac{e^2}{m}~+~4 \pi~(\alpha~+~\beta)~\nu^2~+O(\nu^4),
\label{eq:f}
\end{equation}
here $-\frac{e^2}{m}$, $e$ and $m$ being respectively the proton charge and
mass, is the Thomson scattering amplitude; $\alpha$ is the electric and
$\beta$ the magnetic polarisability of the nucleon. The odd amplitude
$g(\nu)$ is written as
\begin{equation}
4 \pi~g(\nu)~=~-\frac{e^2~\kappa^2}{2~m^2}~\nu~+~4
\pi~\gamma~\nu^3~+O(\nu^5),
\label{eq:g}
\end{equation}
where $\kappa$ is the anomalous magnetic moment of the nucleon and $\gamma$
the forward polarisability.

The next step is made by invoking causality, which, in a time dependent
description of scattering, states that any amplitude $a(t)$ is equal to zero
for $t<0$, that is the scattering wave vanishes before that the incoming
particle collides with the target. This obvious statement has important
consequences on the properties of the scattering amplitudes. In fact,
according to the Titchmarsch theorem \cite{tit}, the following three statements are
equivalent:

1. $a(t)~=~0$ for $t<0$,

2. the Fourier transform $a(\nu)$ is an analytic function,

3. $Re~a(\nu)$ and $Im~a(\nu)$ satisfy the dispersion relations ("Hilbert
transforms"):
\begin{equation}
Re~a(\nu)~=~\frac{1}{\pi}~P~\int_{-\infty}^{\infty} 
\frac{Im~a(\nu)}{(\nu'~-\nu)}~d\nu',
\label{eq:drf}
\end{equation}
\begin{equation}
Im~a(\nu)~=~-~\frac{1}{\pi}~P~\int_{-\infty}^{\infty} 
\frac{Re~a(\nu)}{(\nu'~-\nu)}~d\nu',
\label{eq:drg}
\end{equation}

Before applying Eq.~(\ref{eq:drf}), we have to remind the Optical theorem, a
direct consequence of unitarity, which, in terms of the two amplitudes
$f(\nu)$ and $g(\nu)$ is written as
\begin{equation}
4\pi~Im~f(\nu)~=~\frac{\nu}{2}~(\sigma_{1/2}~+~\sigma_{3/2})~=
~\nu~\sigma_{tot},
\label{eq:optf}
\end{equation}
\begin{equation}
4\pi~Im~g(\nu)~=~\frac{\nu}{2}~(\sigma_{1/2}~-~\sigma_{3/2})~=
~\nu~\Delta\sigma_{tot},
\label{eq:optg}
\end{equation}
where $\sigma_{h}$, with $h=1/2,3/2$, are the helicity dependent cross sections:
$h=1/2$ ($h=3/2$) means that the incident photon and the target nucleon have
antiparallel (parallel) spins.

Now, recalling that $Ref(\nu)$ is even in $\nu$, while $Imf(\nu)$ is odd and
vanishes below the pion threshold $\nu_0$, we can write, for $\nu~<~\nu_0$
\begin{equation}
4\pi~f(\nu)~=~\frac{2}{\pi}~\int_{\nu_0}^{\infty}~d\nu'~\frac{\nu'^2~\sigma_{tot}(\nu')}
{(\nu'^2-\nu^2)}.
\label{eq:drf2}
\end{equation}
The high energy behaviour of the total cross section poses a problem of
convergence of the integral in Eq. (\ref{eq:drf2}), therefore it is convenient to
adopt a subtraction technique
\begin{equation}
4\pi~[f(\nu)~-f(0)]~=~\frac{2}{\pi}~\nu^2~\int_{\nu_0}^{\infty}~d\nu'
~\frac{\sigma_{tot}(\nu')}{(\nu'^2-\nu^2)}.
\label{eq:drf3}
\end{equation}
A similar procedure for the amplitude $g(\nu)$ leads to the unsubtracted dispersion
relation
\begin{equation}
4\pi~g(\nu)~=~\frac{2}{\pi}~\nu~\int_{\nu_0}^{\infty}~d\nu'
~\frac{\nu'~\Delta\sigma(\nu')}{(\nu'^2-\nu^2)}.
\label{eq:drg2}
\end{equation}

Finally, using the LET and the Taylor expansion of the dispersion integrals,
comparing terms of the same order in $\nu$, we arrive at the Baldin sum rule
\cite{bal}
\begin{equation}
\alpha~+\beta~=~\frac{1}{2\pi^2}~\int_{\nu_0}^{\infty}~d\nu'
~\frac{\sigma_{tot}(\nu')}{\nu'^2},  
\label{eq:bal}
\end{equation}
the Gerasimov-Drell-Hearn sum rule \cite{gdh}
\begin{equation}
-\frac{e^2~\kappa^2}{2~m^2}~=~\frac{1}{\pi}~\int_{\nu_0}^{\infty}~d\nu'
~\frac{\sigma_{1/2}(\nu')~-~\sigma_{3/2}(\nu')}{\nu'},  
\label{eq:gdh}
\end{equation}
and the Gell-Mann-Goldberger-Thirring \cite{ggt} formula for the spin polarisability 
\begin{equation}
\gamma~=~\frac{1}{4\pi^2}~\int_{\nu_0}^{\infty}~d\nu'
~\frac{\sigma_{1/2}(\nu')~-~\sigma_{3/2}(\nu')}{\nu'^3},  
\label{eq:ggt}
\end{equation}

\section{The $Q^2$-behaviour}

The GDH integral can be generalised to virtual photon absorption, that is to
electron scattering, by introducing an explicit $Q^2$ dependence. In this way,
 it can be put in relation \cite{ans} with the spin structure function of 
the nucleon. Let us introduce the definition
\begin{equation}
I_1(Q^2)~=~\frac{2m^2}{Q^2}\int_{0}^{1}g_1(x,Q^2)~dx,  
\label{eq:ans}
\end{equation}
where $g_1(x,Q^2)$ is the spin structure function of the nucleon, $Q^2$ the 
momentum transfer and $x~=~\frac{Q^2}{2m\nu}$ is the Bjorken variable. Then
Eq.~(\ref{eq:gdh}) can be rewritten
\begin{equation}
-\frac{2\pi^2\alpha}{m^2}~\kappa^2~=~\int_{\nu_0}^{\infty}~d\nu'
~\frac{\sigma_{1/2}(\nu')~-~\sigma_{3/2}(\nu')}{\nu'}~=
~\frac{8\pi^2\alpha}{m^2}~I_1(0),  
\label{eq:gdh2}
\end{equation}
in this way at the photon point we have
\begin{equation}
I_1(0)~=~-\frac{\kappa^2}{4}.
\label{eq:ans2}
\end{equation}

The extension to non zero $Q^2$ allows in particular a connection of the GDH field to
the sum rules involving the spin structure functions of the nucleon. As examples one
can quote the Ellis-Jaffe sum rule for the proton spin structure \cite{ej}
 \begin{equation}
\Gamma_1^p(Q^2)~=~\int_{0}^{1}g_1^p(x,Q^2)~dx~=~0.185,  
\label{eq:ej}
\end{equation}
whose value turns out to be different from the experimental results. For instance
\cite{smc}, 
\begin{equation}
\Gamma^p_{1~exp}~=~0.141\pm0.011~~~~~~~~~~~~at~~~~Q^2~=~5~(GeV/c)^2,  
\label{eq:ejd}
\end{equation}
as it is well known, this kind of discrepancy has triggered the so call "spin crisis".

Another example of related sum rule is the Bjorken one \cite{bj}
\begin{equation}
\Gamma_1^p~-\Gamma_1^n~=~\frac{1}{6}~\frac{g_A}{g_V}~[1~-~\frac{\alpha_S(Q^2)}{\pi}]~=
~0.191\pm0.002,  
\label{eq:bjo}
\end{equation}
where $\Gamma_1^n$ denotes the integral of the neutron spin structure, $g_A$ and
$g_V$ are the axial-vector and vector constants, respectively, and $\alpha_S(Q^2)$ is
the strong coupling constant. The Bjorken sum rule is in good agreement with the
experimental value of about $0.20$ \cite{smc2}.

There are many ways of generalising the GDH sum rule \cite{lt}, that is of obtaining a
$Q^2$-dependent equation which, at the photon point, reproduces the original Eq.
(\ref{eq:gdh}). One of them uses the integral defined in Eq. (\ref{eq:ans}) 
\begin{equation}
I_1(Q^2)~=~\frac{2m^2}{Q^2}\int_{0}^{1}g_1(x,Q^2)~dx\nonumber
\end{equation}
\begin{equation}
=~\frac{m^2}{8\pi^2\alpha}
\int_{\nu_0}^{i}\frac{1~-x}{1~+y^2}~(\sigma_{1/2}~-~\sigma_{3/2}~-~2~y~\sigma_{LT'})
~\frac{d\nu}{\nu} .
\label{eq:i1}
\end{equation}
Another one introduces the second spin structure function of the nucleon
\begin{equation}
I_2(Q^2)~=~\frac{2m^2}{Q^2}\int_{0}^{1}g_2(x,Q^2)~dx\nonumber
\end{equation}
\begin{equation}
=~\frac{m^2}{8\pi^2\alpha}
\int_{\nu_0}^{i}\frac{1~-x}{1~+y^2}~(\sigma_{3/2}~-~\sigma_{1/2}~-~\frac{2}{\gamma}\sigma_{LT'})
~\frac{d\nu}{\nu} .
\label{eq:i2}
\end{equation}
In both Eqs. (\ref{eq:i1}) and (\ref{eq:i2}) $y=Q/\nu$ and $\sigma_{LT'}$ is the
longitudinal-transverse cross section.

As already mentioned, the integral $I_1(Q^2)$ enters into the Ellis-Jaffe sum rule, while
$I_2(Q^2)$ is involved in the Burkhardt-Cottingham sum rule \cite{bc}
\begin{equation}
I_2(Q^2)~=~\frac{1}{4}~\frac{G_M(Q^2)~-~G_E(Q^2)}{1~+~\frac{Q^2}{4m^2}}
\label{eq:bg}
\end{equation}

There are further generalisations, which make use of integrals involving different combinations
of the cross sections, namely \cite{lt}
\begin{equation}
\frac{\sigma_{1/2}~-~\sigma_{3/2}}{\nu},~~~~~~\frac{\sigma_{1/2}~-
~\sigma_{3/2}}{\nu}
~\frac{1~-~x}{\sqrt{1~+~y^2}},~~~~~~\frac{\sigma_{1/2}~-
~\sigma_{3/2}}{\nu}~(1~-~x).
\end{equation}

\section{The saturation of the GDH sum rule for the nucleon}

The l.h. side of Eq.(\ref{eq:gdh}) is known, since it is determined by the
experimental value of the anomalous magnetic moment of the nucleon; in order to
verify that the r.h. side has the same value one needs a complete knowledge of
the total photoabsorption  cross section as a function of the photon energy. Here
arises the problem of its saturation, since such knowledge is limited by the
availability of experimental data.

Many analyses have been performed in the past, based on the older data. The results
are reported in Table 1 \cite{pp}, where the various evaluations of the GDH-integral
(r.h.side of Eq. (\ref{eq:gdh})) are given separately for the proton and the
neutron, together with their differences, and compared with the experimental value
given by the anomalous magnetic moment.

\begin{table}[th]
\begin{center}
\caption{Different phenomenological analyses of the GDH sum rule.\vspace*{1pt}}
\vspace*{0.5truecm}
{\footnotesize
\begin{tabular}{|c|r|r|r|r|}
\hline
{} &{} &{} &{} \\[-1.5ex]
{} & $I_p^{GDH}~(\mu b)$ & $I_n^{GDH}~(\mu b)$ & $I_{p-n}^{GDH}~(\mu b)$ \\[1ex]
\hline
{} &{} &{} &{} \\[-1.5ex]
GDH sum rule & $204$ & $232$ & $-~28$ \\[1ex]
\hline
\hline
{} &{} &{} &{} \\[-1.5ex]
Karliner \cite{ka} &$261$ &$183$ &$78$ \\[1ex]
Workman-Arndt \cite{wa} &$257$ &$189$ &$68$ \\[1ex]
Burkert-Li \cite{bl} &$203$ &$125$ &$78$ \\[1ex]
Sandorfi et al. \cite{san} &$289$ &$160$ &$130$ \\[1ex]
L.N. Chang et al. \cite{ch} &$294$ &$185$ &$109$\\[1ex]
Drechsel-Krein \cite{dk} &$261$ &$180$ &$81$ \\[1ex]
Bianchi-Thomas \cite{bt} &$207\pm23$ &$226\pm22$ &$-19\pm37$ \\[1ex]
\hline
\end{tabular}\label{tab1} }
\end{center}
\end{table}

\begin{table}[th]
\begin{center}
\caption{Multipole contributions to the photo-excitation of the baryon
resonance in the CQM. $e_i$ and $\sigma_0(i)$ are, respectively, the charge and the
third spin component of the i-th quark, $l_{\lambda~0}$ the third component of the
angular momentum of the third quark.\vspace*{0.5truecm}}  
{\footnotesize
\begin{tabular}{|c|c|}
\hline
{} &{} \\[-1.5ex]
Multipole & Operator \\[1ex]
\hline
{} &{} \\[-1.5ex]
M1/M1 &$e^2_3~\sigma_0(3) +\frac{1}{9}~(2~e^2_3~+~e_1~e_2)~\l_{\lambda~0}$  \\[1ex]
M1/E2 & $0$ \\[1ex]
E2/E2 & $-~\frac{1}{9}~(2~e^2_3~+~e_1~e_2)~\l_{\lambda~0}$ \\[1ex]
E1/E1 & $\frac{2}{3}~e^2_3~\sigma_0(3)~-~\frac{1}{3}~e_1~e_2~[\sigma_0(1)~+
~\sigma_0(2)]~+~\frac{4}{9}~(e^2_3~+e_1~e_2)~\l_{\lambda~0}$ \\[1ex] 
E1/M2 & $-~\frac{2}{3}~e^2_3~\sigma_0(3)~+~\frac{1}{3}~e_1~e_2~[\sigma_0(1)~+
~\sigma_0(2)]~-~\frac{4}{9}~(e^2_3~+e_1~e_2)~\l_{\lambda~0}$ \\[1ex]
M2/M2 & $0$\\[1ex] 
\hline
\end{tabular}\label{tab2} }
\end{center}
\end{table}

All the analyses, except the one by Bianchi and Thomas, overestimate the proton
integral and underestimate the neutron one, leading to wrong (positive) values of the
differences.

The recent measurements of the total photoabsorption cross section of the proton at
MAMI allow the evaluation of the integral directly from the data \cite{pp}. The
resulting value, integrating the experimental data in the interval $200-800$ MeV is
$216\pm6\pm13$ \cite{pp} and seems to be somewhat higher than the expected sum rule.
Further information on this point will be provided by the new data at higher
energy from ELSA (see \cite{pg}). 

The energy range spanned by the present data covers for the major part the proton
resonance region. It is therefore meaningful to try to analyse the GDH sum rule by
means of the Constituent Quark Model, which gives a fairly consistent description of
all resonances. In a simplified, unretarded approach to the photo-excitation of
baryon resonances \cite{dg,mds}, the saturation is substantially
provided by the excitation of the $\Delta$-resonance, as it can be seen in Table 2
\cite{dg}, where the multipole contributions relevant for the considered energy
interval are reported. In fact, summing up all the multipoles, one is left with the
operator $e^2_3~\sigma_0(3)$ coming from the $M1$-multipole, which is mainly due to
the $\Delta$-excitation. 

The sum rule can be explicitely verified using the Isgur-Karl model \cite{ik}. The
anomalous magnetic moment term in the r.h. side of Eq. (\ref{eq:gdh}) can be
evaluated giving
\begin{equation}
\frac{e^2~\kappa^2}{2~m^2}~=~4~-~8~a^2_M~-~10~a^2_D~-~2~\tau_0~a^2_D,  
\label{eq:1gdh}
\end{equation}
where $a_M$ and $a_D$ are the mixing amplitudes of, respectively, the mixed symmetry
and D-wave states.

  If the r.h. side is calculated by averaging the operator
$e^2_3~\sigma_0(3)$ in the nucleon state as given by the Isgur-Karl model one gets a
slightly different result. The equality is recovered \cite{mds} if relativistic
corrections are taken into account, which affect the electric dipole contribution
only.

\section{GDH sum rule with nuclear targets}

The GDH sum rule can be generalized to the case of a nuclear target with mass $M_i$,
charge $Q$, spin $I$ and anomalous magnetic moment $\kappa$ in the following
way \cite{hart}
\begin{equation}
4~\pi^2~\frac{e^2~\kappa^2}{M_i^2}~I~=~\int_{\nu_0}^{\infty}~d\nu
~\frac{\sigma^P(\nu)~-~\sigma^A(\nu)}{\nu}~=~I_{GDH},  
\label{eq:deut}
\end{equation}
where $\sigma^P$ and $\sigma^A$ are the photoabsorption cross sections for parallel
or antiparallel spins, respectively.
The phenomenological value of the r.h. side of Eq. (\ref{eq:deut}) is $0.65~\mu b$.
The theoretical contributions, up to $550~MeV$, of the three channels
$\gamma~d~\rightarrow~n~p$,
$\gamma~d~\rightarrow~d~\piÐ0$, $\gamma~d~\rightarrow~N~N~\pi$ are, respectively,
$-413~\mu b$, $63~\mu b$, $167~\mu b$, which sum up to $I_{GDH}(550)~=~-183~\mu b$.
If, instead of the theoretical value $167~\mu b$ for the
$\gamma~d~\rightarrow~N~N~\pi$ channel, one uses the sum of the corresponding proton
and neutron quantities evaluated by means of a multipole analysis of experimental
data, $331~\mu b$, one gets
$I_{GDH}(550)~=~-19~\mu b$, showing in any case that the higher energy contributions
should be positive and of about the same size \cite{hart}.

A particular attention is being devoted to the deuterium and $^3He$ targets. Besides
being interesting by themselves, such studies are performed with the hope of
extracting information on the neutron photoabsorption cross section. Actually in a
deuteron with $J_z~=~+1$, both proton and neutron have aligned spins if one neglects
the presence of the $D-$state, which has a probability ranging from $4$ to
$7~\%$. In the case of $^3He$, the $S-$state, in the currently accepted
descriptions of the three-nucleon system, has a probability of about $90~\%$: in the
corresponding configuration the proton pair have antiparallel spins, so that the
target polarization is given by the neutron only. Of course, the results have to be
corrected in order to take into account the presence of higher waves (at the $10~\%$
level).

\end{document}